\documentclass{icrc}

\usepackage{times}
 \usepackage{graphicx} 
\begin{document}

\title{Likelihood Analysis of sub-TeV Gamma-rays from RXJ1713-39
with CANGAROO-II}
\author[1]{R. Enomoto}
\affil[1]{Institute for Cosmic Ray Research, University of Tokyo,Kashiwa, 
277-8582 Chiba, Japan}
\author[2]{T. Tanimori}
\affil[2]{Department of Physics, Kyoto University, Sakyo-ku, Kyoto 606-8502, 
Japan}
\author[2]{A. Asahara}
\author[3]{G.V. Bicknell}
\affil[3]{MSSSO, Australian National University, ACT 2611, Australia}
\author[4]{P.G. Edwards}
\affil[4]{Institute of Space and Astronautical Science, Sagamihara, Kanagawa 229-8510, Japan}
\author[5]{S. Gunji}
\affil[5]{Department of Physics, Yamagata University, Yamagata, Yamagata 
990-8560, Japan}
\author[6]{S. Hara}
\affil[6]{Department of Physics, Tokyo Institute of Technology, Meguro-ku, 
Tokyo 152-8551, Japan}
\author[7]{T. Hara}
\affil[7]{Faculty of Management Information, Yamanashi Gakuin University, 
Kofu, Yamanashi 400-8575, Japan}
\author[8]{S. Hayashi}
\affil[8]{Department of Physics, Konan University, Kobe, Hyogo 658-8501,
Japan}
\author[9]{C. Itoh}
\affil[9]{Faculty of Science, Ibaraki University, Mito, Ibaraki 310-8512, Japan}
\author[1]{S. Kabuki}
\author[8]{F. Kajino}
\author[1]{H. Katagiri}
\author[2]{J. Kataoka}
\author[1]{A. Kawachi}
\author[10]{T. Kifune}
\affil[10]{Faculty of Engineering, Shinshu University, Nagano, Nagano 380-8553,
Japan}
\author[2]{H. Kubo}
\author[6]{J. Kushida}
\author[8]{S. Maeda}
\author[8]{A. Maeshiro}
\author[11]{Y. Matsubara}
\affil[11]{STE Laboratory, Nagoya University, Nagoya, Aichi 464-8601, Japan}
\author[12]{Y. Mizumoto}
\affil[12]{National Astronomical Observatory of Japan, Mitaka, Tokyo 181-8588, 
 Japan}
\author[1]{M. Mori}
\author[6]{M. Moriya}
\author[13]{H. Muraishi}
\affil[13]{Ibaraki Prefectural University, Ami, Ibaraki 300-0394, Japan}
\author[11]{Y. Muraki}
\author[7]{T. Naito}
\author[14]{T. Nakase}
\affil[14]{Department of Physics, Tokai University, Hiratsuka, Kanagawa 
259-1292, Japan}
\author[14]{K. Nishijima}
\author[1]{M. Ohishi}
\author[1]{K. Okumura}
\author[15]{J.R. Patterson}
\affil[15]{Department of Physics and Math. Physics, University of Adelaide, 
SA 5005, Australia}
\author[6]{K. Sakurazawa}
\author[1]{R. Suzuki}
\author[15]{D.L. Swaby}
\author[6]{K. Takano}
\author[5]{T. Takano}
\author[5]{F. Tokanai}
\author[1]{K. Tsuchiya}
\author[1]{H. Tsunoo}
\author[14]{K. Uruma}
\author[5]{A. Watanabe}
\author[9]{S. Yanagita}
\author[9]{T. Yoshida}
\author[16]{T. Yoshikoshi}
\affil[16]{Department of Physics, Osaka City University, Osaka, Osaka 558-8585, Japan}

\correspondence{R. Enomoto (enomoto@icrr.u-tokyo.ac.jp)}

\runninghead{R. Enomoto et al: RXJ1713$-$39 (CANGAROO)}
\firstpage{1}
\pubyear{2001}

 \titleheight{15cm} 

\maketitle

\begin{abstract}
We have detected gamma-rays from RXJ1713$-$39 in the energy range between
400 GeV and 5 TeV using a new kind of analysis: likelihood analysis. 
The statistical significance of the measurement
is greater than 8 $\sigma$. 
The details of this analysis method are presented.
\end{abstract}

\section{Introduction}

The CANGAROO experiment is the Imaging Atmospheric Cherenkov Telescope
located in Woomera, South 
Australia. It started with a 3.8-m telescope \citep{kifune1995}.
We now have a 10-m reflector \citep{kawachi2001} and are going to
build a stereo-scopic system \citep{mori2001}.

Supernova remnants are one of the hot topics related to origin of
cosmic rays \citep{yn1999,ellison2000}.
The CANGAROO experiment detected gamma-rays from SN1006 
\citep{tanimori1998a,tanimori2001,hara2001}.
Particle acceleration by a shock wave produced by a supernova
explosion and inverse Compton scattering with micro-wave 
background radiation
\citep{pohl1996,mas1996,masj1996,yy1997,naito1999}
can explain cosmic ray acceleration very well.

RXJ1713$-$39 was found in the ROAST all sky survey \citep{pfeffermann1996},
and was found to have a shell structure.
Hard X-ray emission was observed by ASCA \citep{koyama1997}.
An association with a molecular cloud was found \citep{slane1999}.
The CANGAROO collaboration found an evidence for TeV gamma-ray 
emission from the northwest-rim \citep{muraishi2000}.
The preliminary result of CANGAROO-II (7m) observation in 1999
shows indication of a gamma-ray signal.
In this report, an analysis of RXJ1713$-$39 with CANGAROO-II (10m)
is presented.

\section{Analysis}

\subsection{Data sample}

The observation was carried out during two periods: 23--26 July  and
19--27 August, 2000. 
The pointing direction was as same as that of CANGAROO-I observation
\citep{muraishi2000}, i.e., the NW-rim ($RA$, $\delta$)=($17^h 11^m 56^s.7$,
 $-39^{\circ}31'52''.4(J2000)$), where the X-ray flux is maximum.
The total observation periods were 1419 and 1397 min., 
for
ON- and OFF-source runs, respectively.
We restricted the data to that taken at elevation angles greater
than 60 degrees and without cloud, dew etc., by looking
carefully at the proton shower rate. The resulting good quality data
corresponded to 649 and 642 min. for ON- and OFF-source runs,
respectively.
After removing bad (hot) PMT and bright star hits, we applied a pulse
height cut ($\sim3.3$ photoelectrons) and a timing cut ($\pm40nsec$).
Events with at least one cluster of five-adjacent 
triggered PMTs were then analyzed.

\subsection{Conventional Imaging Cut}

\begin{figure}[t]
 \vspace*{2.0mm} 
 \includegraphics[width=8.3cm]{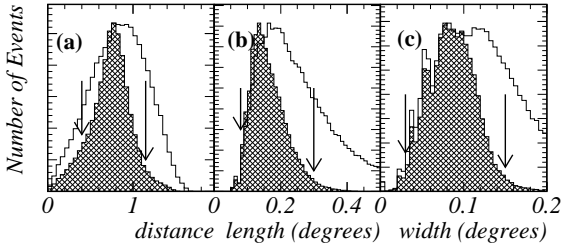} 
 \caption{Shape parameter distributions: (a) $distance$, (b) 
$length$, and (c) $width$. The blank histograms were obtained from
the OFF-source run. The hatched histograms are Monte-Carlo gamma-ray events.
The cut positions are indicated by arrows.
 }
 \label{fig:cut}
 \end{figure}

First we rejected energetic multi-cluster events as follows.
We calculated the energy-ratio ($E_{other}/E_{max}$) 
between the most energetic cluster ($E_{max}$)
and the others ($E_{other}$). 
Because the gamma-ray events most likely have a single cluster,
we rejected events with this ratio greater than 25\%.
Then, the event shape parameters \citep{hillas1985,weekes1989} 
of $distance$, $length$, $width$, $asymmetry$ and $\alpha$ were calculated
for the most energetic clusters (Fig \ref{fig:cut}).
The blank histograms are obtained from
the OFF-source run. The hatched histograms are Monte-Carlo gamma-ray events.
The cut positions are indicated by the arrows.
Here, we refer to this as a $square~cut~analysis$.
The cut dependences of the $\alpha$ distributions are shown in Fig
\ref{fig:square}: 
 \begin{figure}[hbt]
 \vspace*{2.0mm} 
 \includegraphics[width=8.3cm, height=14.0cm]{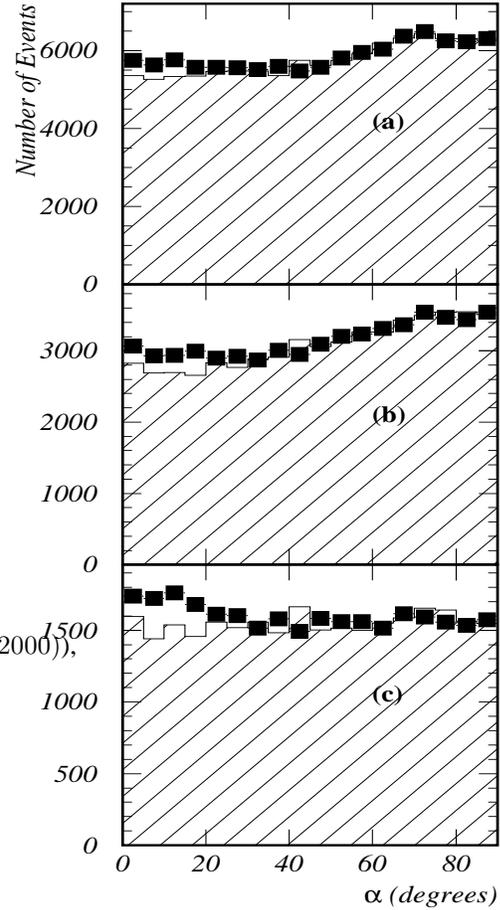} 
 \caption{Image orientation angle ($\alpha$) distributions for ON-source data
(solid squares) and OFF (hatched histogram), for a ``square cut''
analysis; (a) no cut, (b) $distance$ cut, and (c) ($distance$,
$length$, and $width$) cut. }
 \label{fig:square}
 \end{figure}
Fig \ref{fig:square}-(a) was obtained without any shape cuts, and
(b) was obtained by a $distance$ cut. By adding $length$ and $width$ cuts,
we obtained Fig \ref{fig:square}-(c). 
The data points with statistical error bars were obtained by ON-source runs,
and the hatched histogram by OFF-source runs, respectively.
Better signal-to-noise ratios (S/N-ratios)
were obtained by tighter cuts. 
The normalization of OFF events was done using the number of events
with $\alpha>25$ degrees. This agreed with the time-interval
ratios between the ON- and OFF-source runs.
The obtained signal levels were:
(a) $1510\pm 234 (6.4\sigma)$, (b) $1129\pm 169 (6.7\sigma)$, and
(c) $931\pm 127 (7.3\sigma)$, respectively.

\subsection{Likelihood}

\begin{figure}[t]
 \vspace*{2.0mm} 
 \includegraphics[width=8.3cm]{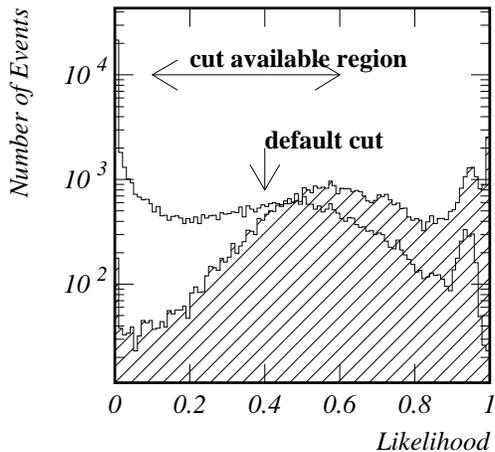} 
 \caption{Likelihood distributions for Monte-Carlo gamma rays (the hatched
region) and OFF-source events (the blank histogram). }
 \label{fig:lval}
 \end{figure}

\begin{figure}[t]
 \vspace*{2.0mm} 
 \includegraphics[width=8.3cm]{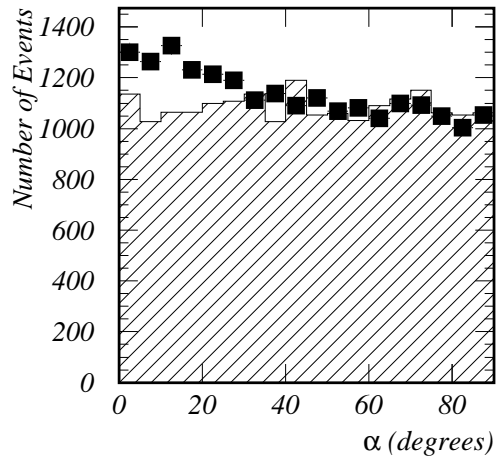} 
 \caption{Image orientation angle ($\alpha$) distributions for ON-source run
(data points with error bars) and OFF (hatched histogram) for a likelihood cut
analysis. }
 \label{fig:likelihood}
 \end{figure}

It is well known that there are energy dependences in the standard shape 
parameters \citep{hillas1998}. One way to minimize these
effects is a likelihood analysis \citep{enomoto2001}.
Before doing this, we have observed a strong
correlation between the $distance$ and the other shape parameters.
This is considered to be due to the edge effect of the focal plane
detector (i.e., camera). We, therefore, need a $distance$ cut before
starting a likelihood analysis. Because the $distance$ also has a small
energy dependence, we corrected it by a linear function
of energy.
After applying an energy-dependent $distance$ cut, we calculated 
the likelihood using the above shape parameters ($length$, $width$,
and $asymmetry$)
for both gamma-ray and proton assumptions. 
In order to estimate the gamma-ray probability-density-function (PDF), 
we used Monte-Carlo
gamma-ray events; for protons, the OFF-source events were
used. 
In the Monte-Carlo simulation, gamma-rays
with a Crab-like spectrum ($E^{-2.5}$) were generated.
In practice, we made two-dimensional
histograms, for example, $log(ADC)$ vs $length$.
We assumed that $ADC$ is proportional to the energy
of the showers.
The total number of events was normalized to unity
and two-dimensional (2D) PDFs were
obtained.
We defined the likelihood-ratio ($L$) as
$$L=Prob(\gamma)/(Prob(\gamma)+Prob(p))$$,
\noindent where, $Prob$ means
product of 2D-PDFs of each shape parameter for the gamma-ray or
proton assumption.
Here, a one-to-one contamination of gamma-rays to protons was
assumed (this can be modified in the future), in spite of the fact that
protons are dominant in cosmic-rays.
The distributions of $L$ are shown in Fig \ref{fig:lval}.
The hatched histogram was $L$ for Monte-Carlo gamma-rays and the blank
OFF-source events.
The final event samples
were obtained by cutting events with $L>0.4$, which is indicated
by the arrow in Fig \ref{fig:lval}.
The resulting $\alpha$ distribution is plotted in Fig. \ref{fig:likelihood}.
The data points with statistical error bars were obtained by ON-source run,
and the hatched histogram by OFF-source run, respectively.
The number of excess events was 946$\pm$108($8.7\sigma$).
In order to check the algorithm, we tried changing the background sample,
i.e., from OFF- to ON-source events to make proton PDFs.
The number of excess events changed to 938$\pm$108($8.7\sigma$),
allowing us to conclude that there are no event-specific biases.
The S/N-ratio was greatly improved by this analysis.
The energy threshold for this analysis was estimated to be $~400$ GeV
with the present CANGAROO-II system.

The $\alpha$ distribution is wider than that for a point source assumption
(typically within $15^{\circ}$), consistent with the
previous observation \citep{muraishi2000}.
Roughly speaking, the observed flux agrees with the previous observation
\citep{muraishi2000}. A more precise flux calculation is now being
carried out,
based on the following systematic error studies.

%


\section{Possible Systematics}

Here, we discuss the possible systematic errors in the acceptance
calculation using this likelihood method.
The acceptance was calculated using the Monte-Carlo 
method \citep{enomoto2001}. 
Gamma-rays
with a Crab-like spectrum ($E^{-2.5}$) were generated.
The measured values of the detector parameters, such as
the point spread function (PSF) of the mirrors, the reflectivity, etc.
we used. 
In order to check, we analyzed Crab data 
(observed in December 2000 with the 10-m telescope)
and compared the obtained flux
with previous measurements \citep{tanimori1998b,aharonian2000}.
They are consistent with each other within 12\% at $\sim 3$ TeV.
On the other hand, the statistical error of the Crab measurement was 17\%.
In order to check cut dependence of the analysis,
we also changed the cut value of $L$, the clustering methods,
and the threshold, and obtained the fluxes each time.
From the above procedures, we estimated a systematic uncertainty of
15.4\%. The stable region of $L$ in the acceptance calculation
is shown in Fig \ref{fig:lval} by the horizontal arrow. The
acceptance is considered to be stable within 6.5\% in this
region with respect to the $L$ cut.

\subsection{Energy Spectrum}

For the energy scale ambiguities, we considered the error on mirror
reflectivity, PSFs, the effects of Mie scatterings,
and the ambiguity on the single-photon pulse height. 
The estimated value was 20\%.

\subsection{Angular Resolution}

The pointing error of the telescope system can be checked by looking at
bright stars in various observation periods. To date, we have
only had bright stars beyond one-degree from the centre of the field of view
where there were significant edge effects and aberrations.
For now we have adopted a systematic error of 0.1 degree. 

\section{Discussion}

An improvement in the S/N-ratio was clearly demonstrated so far. 
Generally speaking, the gamma-ray signal increased with the background
remaining the same, when compared with the traditional square cut.

As for the $asymmetry$ parameter, only a positive or negative
cut can be applied in the $square~cut$ $analysis$, greatly
reducing the number of events accepted. However, 
a likelihood analysis can save
this situation. Almost all of the parameters can be put into the
PDFs, even if there are small differences between gamma-rays and protons.
It is necessary, however, to be careful of the dependencies between
the various parameters.

The most important thing is that this kind of analysis can reduce 
any ``human bias", because the number of cut parameters is very small.
The systematic error, therefore, should be smaller than the
$square~cut~analysis$.
The small number of cut parameters makes it easy to  estimate
the systematic errors.

Usually, the cut value of $L$ should be around $\sim 0.5$; allowing 
automated analysis for any situations such as
different elevation angles, etc.
For the analysis of stereoscopic observations \citep{enomoto2001},
the product of $L$ of many telescopes can be used as a 
single-cut parameter.
In this case it is necessary to tune the Monte-Carlo simulations as 
accurately as possible.

\section{Conclusion}

We have measured gamma-rays from RXJ1713$-$39 in the energy range between
400\, GeV and 5\,TeV. The statistical significance of the measurement
is greater than 8\,$\sigma$. 
We have used the new likelihood method of analysis.
We obtained a better signal-to-noise ratio compared with the standard
analysis. Also we confirmed the systematic error of this analysis
is sufficiently small.

\begin{acknowledgements}

This work was supported by the Center of Excellence,
and a
Grant-in-Aid for Scientific
Research by the Japan Ministry of Education, Science, Sports and
Culture.
Also it was supported by Australian Research Council.

\end{acknowledgements}

\end{document}